\documentclass[letterpaper,twocolumn,showpacs,prl,aps]{revtex4-1}

\usepackage{amsmath}  
\usepackage{amsfonts} 
\usepackage{graphicx} 
\usepackage{amssymb}

\begin{document}

\title{Steady flows, nonlinear gravitostatic waves and  Zeldovich pancakes in a Newtonian gas}

\author{Eugene B. Kolomeisky}

\affiliation
{Department of Physics, University of Virginia, P. O. Box 400714, Charlottesville, Virginia 22904-4714, USA}

\date{\today}

\begin{abstract}
We show that equations of Newtonian hydrodynamics and gravity describing one-dimensional steady gas flow possess nonlinear periodic solutions.  In the case of a zero-pressure gas the solution exhibits hydrodynamic similarity and is universal:  it is a lattice of integrable density singularities coinciding with maxima of the gravitational potential.  With finite pressure effects included, there exists critical matter density that separates two regimes of behavior.  If the average density is below the critical, the solution is a density wave which is in phase with the wave of the gravitational potential.  If the average density is above the critical, the waves of the density and potential are out of phase.  Traveling plane gravitostatic waves are also predicted and their properties elucidated.  Specifically, subsonic wave is made out of two out of phase oscillations of matter density and gravitational potential.  If the wave is supersonic, the density-potential oscillations are in phase.   
\end{abstract}


\maketitle

\textit{Introduction}. - It is well-known that attractive long-range nature of gravity is incompatible with static uniform distribution of matter:  the ensuing gravitational instability is the key to understanding the large-scale structure of the Universe \cite{ZN}.  While linear stage of the instability is well-understood, what happens beyond it continues to be a subject of inquiry.  In a pioneering study \cite{Z} Zeldovich put forward an approximate solution to the nonlinear problem according to which initial anisotropic density fluctuation of zero-pressure matter first condenses into a wall-like object, an integrable density singularity, dubbed the "Zeldovich pancake".  Further studies \cite{ZS} confirmed this idea also pointing to the existence of other collapsed objects such as filaments and compact clumps of matter eventually leading to current understanding of the cosmic web as an hierarchy of condensed structures \cite{web}.   

While anisotropic interim structures further fragment and collapse under their own gravity, a question arises whether noncondensing counterparts of these objects can exist.  Below we carry out classification of one-dimensional steady flows and a related problem of traveling waves in a Newtonian gravitating gas, and in the zero-pressure case indeed discover a universal structure that may be viewed as a lattice of Zeldovich pancakes - static in the case of the steady flow or traveling as a wave.  Additionally, accounting for the effects of pressure, we find situations with nearly uniform distribution of matter which, gravity notwithstanding, is unexpected.  While strictly one-dimensional setup is idealized, usefulness of the analysis stems from two observations:\newline
(i) nonlinear effects are treated exactly which is a benefit of one-dimensionality; \newline
(ii) traveling wave solutions that we predict are counterparts of plane waves of electromagnetism and elasticity theory which are of general importance even though they are also one-dimensional.  

Our study is modeled after an investigation of nonlinear plasma waves in an electron gas \cite{AL,EBK}.  Indeed the two settings are similar as both the Coulomb and gravitational interactions fall as an inverse separation between the particles.  The important  difference is that gravitational interaction is attractive while the Coulomb interaction can be either repulsive or attractive depending on whether interacting particles carry like or unlike charges.  
   
Our starting point is standard \cite{ZN}, the system of equations of Newtonian hydrodynamics and gravity for an ideal fluid described by the local position- and time-dependent mass density $\rho(\textbf{r},t)$ and velocity $\textbf{v}(\textbf{r},t)$ fields, which are related by the continuity equation
\begin{equation}
\label{continuity}
\frac{\partial \rho}{\partial t}+\nabla \cdot(\rho\textbf{v})=0
\end{equation}
The equation of motion of the fluid is given by the Euler equation of hydrodynamics 
\begin{equation}
\label{2nd_law}
\frac{\partial \textbf{v}}{\partial t} +(\textbf{v}\cdot \nabla)\textbf{v} =-\nabla \left [w(\rho) + \phi\right ]
\end{equation}
where $w(\rho)$ is the heat function per unit mass of fluid and $\phi$ is the gravitational potential determined by the density $\rho$ via the Poisson equation
\begin{equation}
\label{Poisson}
\nabla^{2}\phi=4\pi G\rho
\end{equation}
where $G$ is the universal gravitational constant.  The heat function $w(\rho)$ is related to the pressure $p$ as $\nabla w=\nabla p/\rho$.  According to the Euler equation (\ref{2nd_law}), the fluid is accelerated both by the gradient of the heat function (pressure) and the gravitational field $-\nabla \phi$:  the latter promotes the collapse while the former opposes it.  Hereafter we assume that there exists fixed average matter density $\overline{\rho}$ and this will be confirmed by an explicit calculation.    

\begin{figure}
\includegraphics[width=1.0\columnwidth, keepaspectratio]{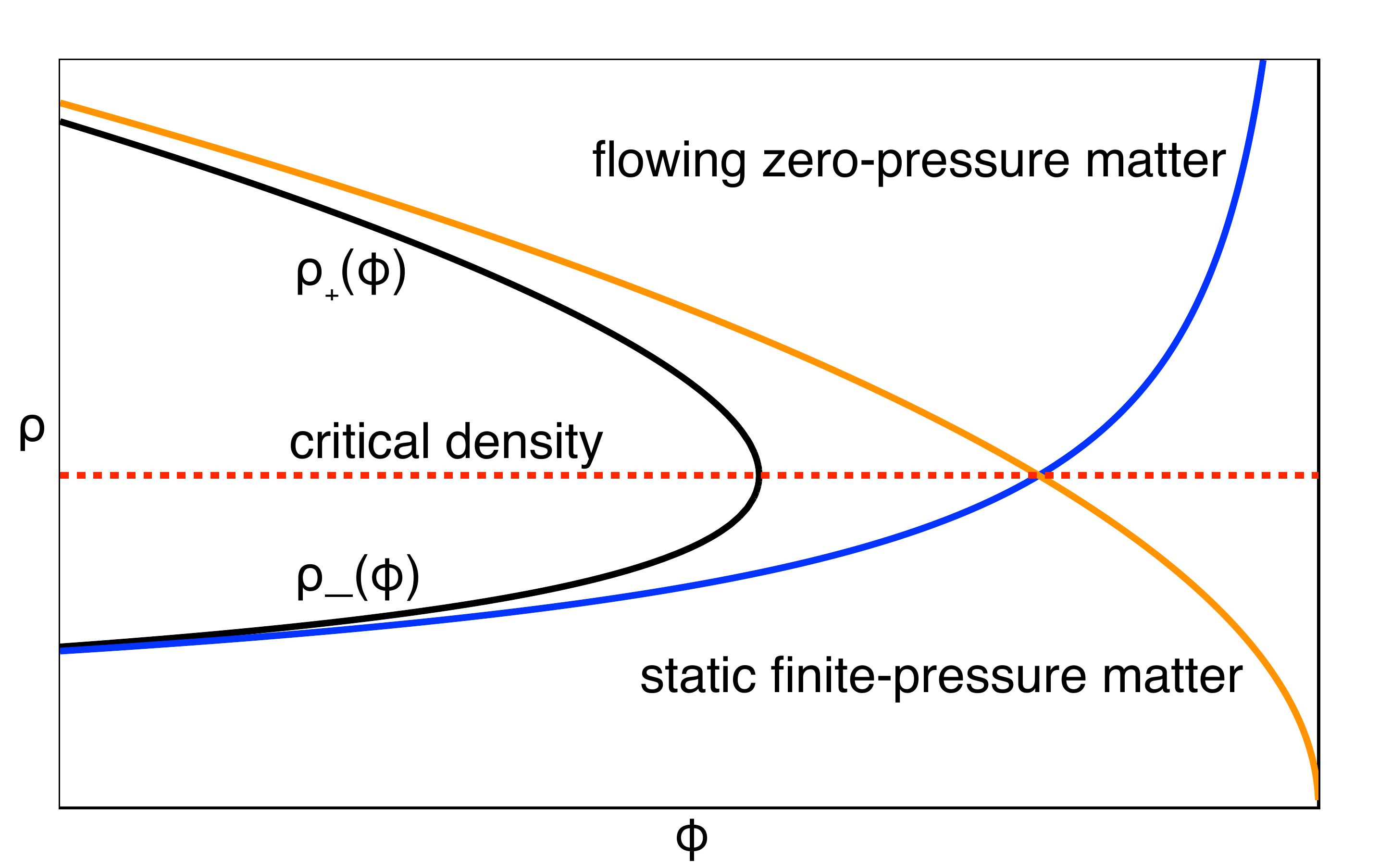} 
\caption{(Color online) Sketches of solutions $\rho(\phi)$ to Eq.(\ref{combined}).  In generic case (black) there are two branches, $\rho_{+}(\phi)$ and $\rho_{-}(\phi)$, separated by the critical density $\rho_{c}$ (\ref{critical_density}) indicated by straight dotted red line.  For flowing zero-pressure matter, $w=0$, the solution $\rho_{-}(\phi)$, Eq.(\ref{zero_pressure_PE}), is shown in blue.  For static, $j=0$, finite-pressure matter corresponding solution $\rho_{+}(\phi)$ is sketched in orange.}
\label{branches}
\end{figure}

\textit{One-dimensional steady flow of Newtonian gravitating gas}. - We begin with the case of a one-dimensional motion along the $x$ axis, $\textbf{v}=(v,0,0)$, and seek solutions for the density $\rho$, velocity $v$ and gravitational potential $\phi$ that depend only on $x$.  Then Eqs.(\ref{continuity})-(\ref{Poisson}) transform into 
\begin{equation}
\label{1d_continuity}
(\rho v)'=0
\end{equation}
\begin{equation}
\label{1d_Euler}
vv'=-(w+\phi)'
\end{equation}
\begin{equation}
\label{1d_Poisson}
\varphi''=4\pi G\rho
\end{equation}
where the prime is a shorthand for the derivative with respect to $x$.  Integrating Eq.(\ref{1d_continuity}) we find
\begin{equation}
\label{1d_continuity_integrated}
j=\rho v
\end{equation}
where $j$ is a constant mass flux;  hereafter we assume that $j\geqslant0$, i.e. the matter flows along the positive $x$ direction.  Integrating Eq.(\ref{1d_Euler}) we arrive at the Bernoulli equation
\begin{equation}
\label{Bernoulli}
\frac{v^{2}}{2}+w+\phi=const,
\end{equation}
the law of conservation of energy for a particle of unit mass and internal energy $w$ in potential energy landscape $\phi(x)$.  Eqs.(\ref{1d_continuity_integrated}) and (\ref{Bernoulli}) can be combined into an expression
\begin{equation}
\label{combined}
-\phi + const=\frac{j^{2}}{2\rho^{2}}+w(\rho)
\end{equation}
which can be inverted to infer the dependence of the density on the potential $\rho(\phi)$ that appears in the right-hand side of the Poisson equation (\ref{1d_Poisson}).  Since the heat function $w$ is a monotonically increasing function of the density $\rho$, the right-hand side of Eq.(\ref{combined}) has a minimum at a critical density $\rho_{c}$ that is a solution to the equation
\begin{equation}
\label{critical_density}
j=\rho_{c}^{3/2}\left (\frac{dw}{d\rho}\right )_{\rho=\rho_{c}}^{1/2}.
\end{equation} 
There are two branches of the function $\rho(\phi)$ labeled $\rho_{+}(\phi)$ and $\rho_{-}(\phi)$ and sketched in Figure \ref{branches} meeting at $\rho=\rho_{c}$, and two classes of solutions to Eqs.(\ref{1d_Poisson}) and (\ref{combined}).  Because the right-hand side of Eq.(\ref{combined}) is non-negative, there is an upper bound to the gravitational potential $\phi$  hereafter set at zero, i.e. $\phi \leqslant 0$.

\begin{figure}
\includegraphics[width=1.0\columnwidth, keepaspectratio]{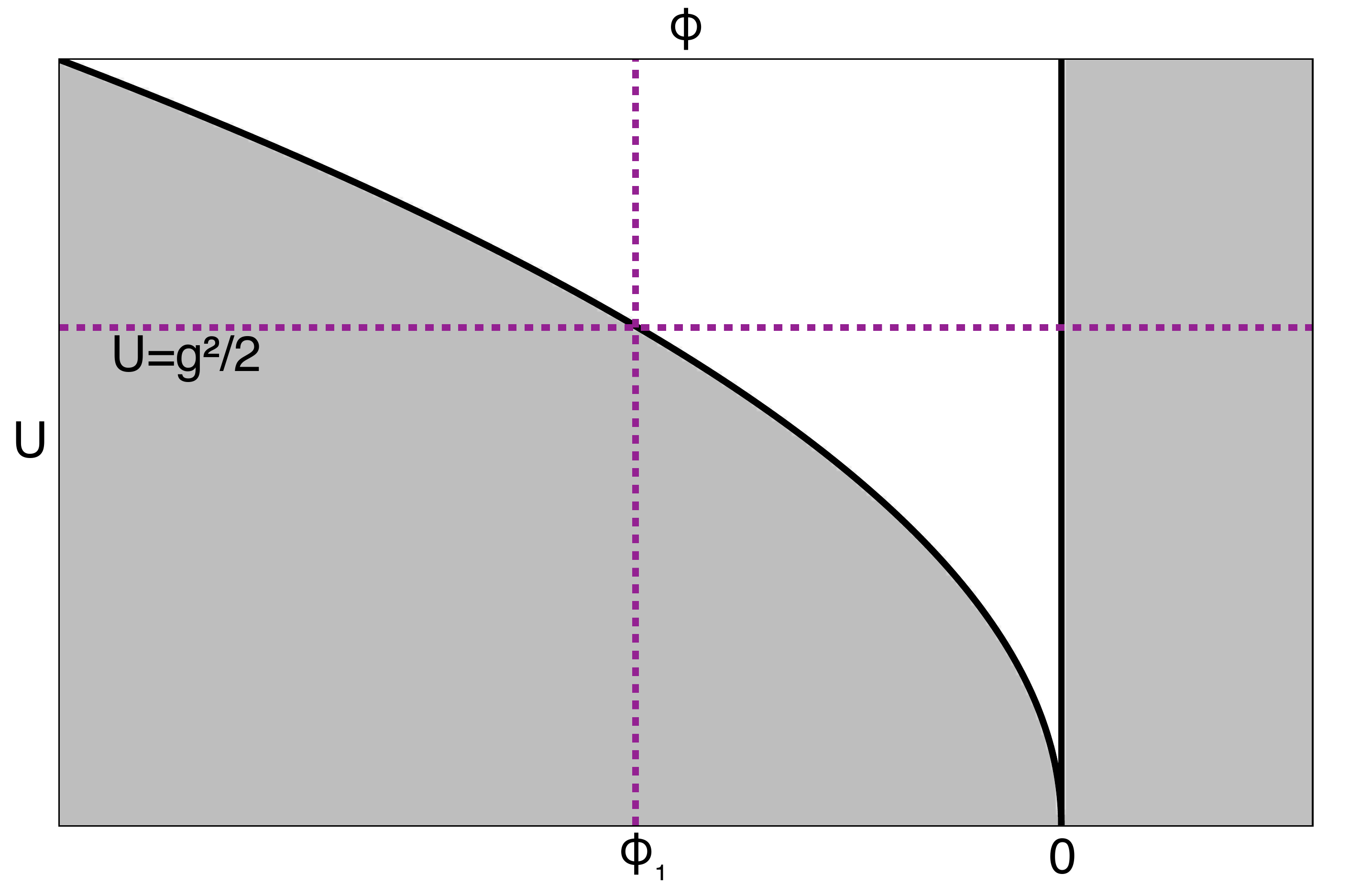} 
\caption{(Color online) Sketch of potential well (for fictitious particle of unit mass and position $\phi$) formed by monotonically decreasing potential energy function $U(\phi)$, Eq.(\ref{energy_integral}), and the constraint $\phi\leqslant 0$.  For fixed energy $g^{2}/2$ the motion is oscillatory:  at leftmost turning point $\phi=\phi_{1}$ the particle temporarily stops and turns around while at rightmost point $\phi=0$ it bounces off the wall and just reverses its direction of motion.  Grayscale regions are not accessible for motion.}
\label{well}
\end{figure}

If $\phi$ is viewed as a position of a fictitious classical particle of unit mass, $x$ as a time, and $4\pi G \rho(\phi)$ as an exerted force, Eq.(\ref{1d_Poisson}) parallels Newton's second law of motion for the particle.  The first integral of (\ref{1d_Poisson}) has the form
\begin{equation}
\label{energy_integral}
\frac{\phi'^{2}}{2}+U(\phi)=\frac{g^{2}}{2},~~~U(\phi)=-4\pi G \int_{0}^{\phi}\rho(\phi)d\phi
\end{equation}  
where $U(\phi)$ is the "potential energy", and the integration constant $g^{2}/2$ is the "energy".  Solutions of Eqs.(\ref{1d_Poisson}) and (\ref{combined}) can be classified according to possible motions of a classical particle in the field of potential energy $U(\phi)$ for different values of the energy $g^{2}/2$.  An example pertinent to our study is sketched in Figure \ref{well}.  Integrating the first-order differential equation (\ref{energy_integral}) we find 
\begin{equation}
\label{general_solution}
x= \pm\int_{0}^{\phi} \frac{d\phi}{\sqrt{g^{2}-2U(\phi)}}
\end{equation}   
Since the motion of the particle in the field of potential energy $U(\phi)$ is limited by the condition $\phi\leqslant0$, the gravitational potential exhibits slope discontinuity at $\phi=0$.  The parameter $g$ in (\ref{energy_integral}) is then the magnitude of the gravitational field on both sides of $\phi=0$, i.e. $\lim_{x\to \pm 0}\phi'(x)=\mp g$.    Likewise, the density $\rho$ is a non-analytic function of the gravitational potential at $\phi=0$ where it is given by the critical density $\rho=\rho_{c}$, Eq.(\ref{critical_density}).  

If $U(\phi<0)$ is monotonically decreasing, as sketched in Figure \ref{well}, the particle motion is finite.  The gravitational potential $\phi$ is then a periodic function, $\phi(x+\lambda)=\phi(x)$, varying between its minimal value, $\phi=\phi_{1}<0$, the root of the equation $U(\phi)=g^{2}/2$, and its maximal value, $\phi=0$.  The period $\lambda$ is given by 
\begin{equation}
\label{period}
\lambda=2\int_{\phi_{1}}^{0}\frac{d\phi}{\sqrt{g^{2}-2U(\phi)}}.
\end{equation} 
The matter density $\rho(x)$ is also periodic with the same period.  The density at $\phi=\phi_{1}$ is $\rho=\rho_{1}=\rho(\phi_{1})$ while at $\phi=0$ it is $\rho=\rho_{c}$.  A relationship between $\rho_{1}$ and $\rho_{c}$ depends on the branch of the function $\rho(\phi)$, Eq.(\ref{combined}), singled out by the average density
\begin{equation}
\label{avdensity}
\overline{\rho}=\frac{2}{\lambda}\int_{-\lambda/2}^{0}\rho(x)dx=\frac{g}{2\pi G\lambda}
\end{equation}
where we employed the Poisson equation (\ref{1d_Poisson}).  Eq.(\ref{avdensity}) relates the average density $\overline{\rho}$, the amplitude of the gravitational field $g$, and the mass flux $j$, thus implying that only two of these three parameters are independent.      

We already observed, Figure \ref{branches}, that there are two branches of the function $\rho(\phi)$ given by Eq.(\ref{combined}), $\rho_{\pm}(\phi)$ ($\rho_{-}\leqslant \rho_{+}$), coalescing at $\rho=\rho_{c}$.  The lower, $\rho_{-}(\phi)\leqslant\rho_{c}$, is a monotonically increasing function of its argument while the upper, $\rho_{+}(\phi)\geqslant \rho_{c}$, is monotonically decreasing.  Tuning the average density $\overline{\rho}$ (\ref{avdensity}) relative to the critical density $\rho_{c}$ (\ref{critical_density}) singles out one of the two possible classes of solutions to Eqs.(\ref{1d_Poisson}) and (\ref{combined}).  Specifically, if $\overline{\rho}<\rho_{c}$, the $\rho_{-}(\phi)$ branch is selected.  The density profile $\rho_{-}(x)$ is then a wave oscillating between $\rho=\rho_{1}$ and $\rho=\rho_{c}>\rho_{1}$;  density maxima $\rho=\rho_{c}$ coincide with maxima of the gravitational potential $\phi=0$.  If, on the other hand, $\overline{\rho}>\rho_{c}$, the $\rho_{+}(\phi)$ branch is singled out.  The density profile $\rho_{+}(x)$ is a wave oscillating between $\rho=\rho_{1}$ and $\rho=\rho_{c}<\rho_{1}$;  density minima $\rho=\rho_{c}$ now coincide with maxima of the potential $\phi=0$.  

To expand on this reasoning, we begin with the case of a zero-pressure matter, $w(\rho)=0$ which may be viewed as mimicking dark matter.  Now the critical density $\rho_{c}$, Eq.(\ref{critical_density}), is infinite, and Eq.(\ref{combined}) has a single solution sketched in blue in Figure \ref{branches} and given by 
\begin{equation}
\label{zero_pressure_PE}
\rho_{-}(\phi)=\frac{j}{\sqrt{-2\phi_{-}}},~~~U(\phi_{-})=4\pi Gj\sqrt{-2\phi_{-}}
\end{equation}
where we also included corresponding potential energy function $U(\phi_{-})$, Eq.(\ref{energy_integral}).  The function $U(\phi_{-})$ subject to the constraint $\phi_{-}\leqslant 0$ is a potential well as sketched in Figure \ref{well}, and the left turning point of motion of the fictitious particle is $\phi_{1}=-g^{4}/128\pi^{2}G^{2}j^{2}$.  The function $U(\phi_{-})$ is homogeneous of the degree $1/2$ thus implying that the particle motion exhibits the property of mechanical similarity \cite{LL1}.   This translates into hydrodynamic similarity of the steady flow that is highlighted by adopting the following units:
\begin{equation}
\label{units}
[x]=\frac{g^{3}}{128\pi^{2}G^{2}j^{2}},~[\phi]=\frac{g^{4}}{128\pi^{2}G^{2}j^{2}},~[\rho]=\frac{8\pi Gj^{2}}{g^{2}} 
\end{equation}
where $[z]$ stands for the unit of $z$.  With these choices the Poisson equation (\ref{1d_Poisson}) acquires the form $\phi_{-}''=\rho_{-}/4$, $\rho_{-}=1/\sqrt{-\phi_{-}}$, there is unit mass flux flowing across the system, and the problem is parameter free.  Then performing the integrations in Eqs.(\ref{general_solution}) and (\ref{period}) we find 
\begin{equation}
\label{zero_pressure_potential}
\pm \frac{3}{4}x=2-3\left (1-\sqrt{-\phi_{-}}\right )^{1/2}+\left (1-\sqrt{-\phi_{-}}\right )^{3/2},
\end{equation}
and $\lambda=16/3$.  The universal function $\phi_{-}(x)$ obtained by inverting Eq.(\ref{zero_pressure_potential}) gives the gravitational potential within its period $[-8/3,8/3]$, and has to be periodically continued beyond this interval.  The function $\phi_{-}(x)$ has a maximum at $x=0$, $\phi_{-}(x\rightarrow 0)=-|x|$, and minima at $x=\pm 8/3$, $\phi_{-}(x\rightarrow \mp 8/3)= -1+(x\pm 8/3)^{2}/8$.  An unbounded relative of the solution (\ref{zero_pressure_potential}) has been mentioned previously \cite{AL} in the context of a zero-pressure electron gas without compensating positive charge background.  

\begin{figure}
\includegraphics[width=1.0\columnwidth, keepaspectratio]{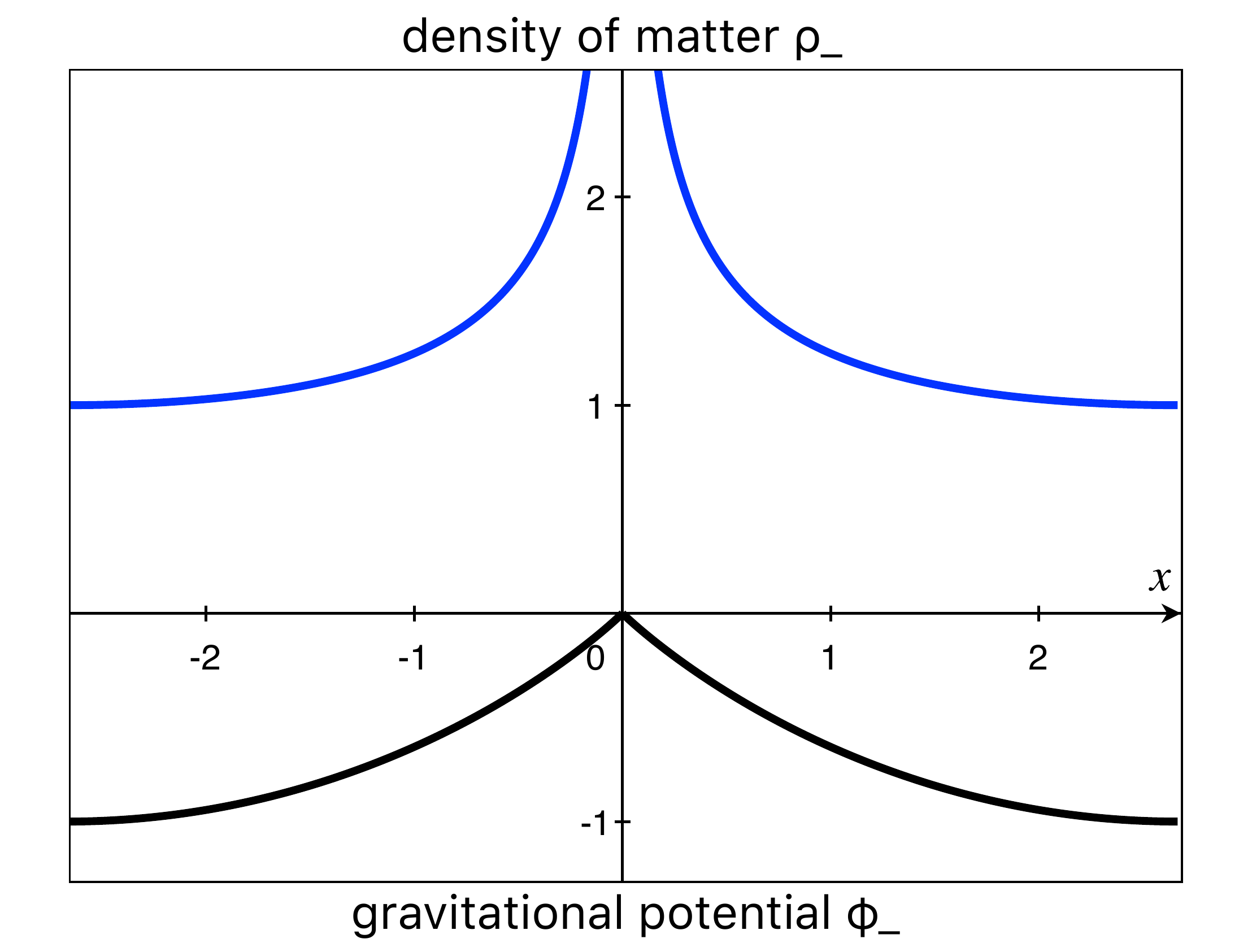} 
\caption{(Color online) One period of spatial variation of the gravitational potential $\phi_{-}$ (black) (\ref{zero_pressure_potential}), and density $\rho_{-}=1/\sqrt{-\phi_{-}}$ (blue) in reduced units (\ref{units}) for a zero-pressure matter.  Density singularity at the origin is the Zeldovich pancake.}
\label{zero}
\end{figure}

The gravitational potential $\phi_{-}(x)$ and corresponding density of matter $\rho_{-}=1/\sqrt{-\phi_{-}}$ are plotted in Figure \ref{zero}.
The density is smallest at the minima of the gravitational potential, $\rho_{-}(x=\pm 8/3)=1$, and singular at its maxima, $\rho_{-}(x\rightarrow 0)=1/\sqrt{|x|}$.  This fact, however, poses no conceptual difficulty as the singularity is integrable.  Indeed the average density of matter according to Eq.(\ref{avdensity}) is $\overline{\rho}=3/2$. In the original physical units (see Eqs.(\ref{units})) this translates into the equalities 
\begin{equation}
\label{equalities}
g=2\sqrt{3\pi G}\frac{j}{\overline{\rho}^{1/2}},~~~\lambda=\sqrt{\frac{3}{\pi G}}\frac{j}{\overline{\rho}^{3/2}}
\end{equation}
supplying us with dependences of the amplitude of the gravitational field $g$ and the period of the structure $\lambda$ on parameters of the problem $\overline{\rho}$ and $j$.  Curiously, for $j$ fixed both $g$ and $\lambda$ diverge with decrease of the average density as $\overline{\rho}^{-1/2}$ and $\overline{\rho}^{-3/2}$, respectively.

Physical reason for the density singularity at the maxima of the gravitational potential can be understood with the help of the law of conservation of energy (\ref{Bernoulli}) which in reduced units (\ref{units}) has the form  $v^{2}+\phi=0$ (unit of the velocity is $g^{2}/8\pi Gj$).  Indeed, a particle starting at the minimum of the potential energy landscape $\phi=-1$ has unit velocity;  it then arrives at the maximum $\phi=0$ with zero velocity.  Thus particle accumulation at the maxima of $\phi$ is the reason why the density is singular.  This mechanism essentially describes the onset of wave breaking \cite{breaking}.  Density singularities localized at the maxima of the gravitational potential are steady counterparts of the Zeldovich pancakes.  Like original Zeldovich pancakes \cite{Z,ZS}, they are caustics of the density field.

Unlimited accumulation of the particles at the maxima of the gravitational potential, artifact of the zero-pressure case, is halted by finite-pressure effects:  the critical density $\rho_{c}$ (\ref{critical_density}) replaces infinite density at the center of the Zeldovich pancake, Figure \ref{zero}.  As the average density $\overline{\rho}$ increases approaching the critical density $\rho_{c}$ (\ref{critical_density}), the amplitude of the density wave steadily decreases vanishing as $\overline{\rho}\to \rho_{c}-0$, and this class of solutions disappears.  

If $\overline{\rho}>\rho_{c}$, upper branch of Eq.(\ref{combined}), $\rho_{+}(\phi)\geqslant \rho_{c}$, governs the character of the solutions to the Poisson equation (\ref{1d_Poisson}).  Here the $j=0$ limit plays a role similar to that of the zero-pressure case for the $\overline{\rho}<\rho_{c}$ domain.  Now the critical density $\rho_{c}$ (\ref{critical_density}) vanishes, and the Bernoulli equation (\ref{Bernoulli}) turns into the equation of hydrostatics, $w(\rho)+\phi=0$.  Its single solution, $\rho_{+}(\phi)$, sketched in Figure \ref{branches} in orange, a monotonically decreasing function vanishing at $\phi=0$, leads to the potential energy function $U(\phi_{+}\leqslant 0)$ in Eq.(\ref{energy_integral}) that is again a potential well for the fictitious particle.  Thus the gravitational potential $\phi_{+}(x)$ is a periodic function varying between $\phi_{1}<0$ and zero.  Similarly, the density is a periodic function of position.  However the difference from the $\overline{\rho}<\rho_{c}$ case is that now the density is largest at the minima of the gravitational potential, and smallest at its maxima $\phi=0$ where $\rho_{+}=0$.  Restoring $j$ to a finite value also makes $\rho_{c}$ (\ref{critical_density}) nonzero thus lifting the density minimum to $\rho_{+}=\rho_{c}$.  As the average density $\overline{\rho}$ decreases approaching the critical density $\rho_{c}$ (\ref{critical_density}), the amplitude of the density wave steadily decreases vanishing as $\overline{\rho}\to \rho_{c}+0$, and this class of solutions disappears.

Even though density waves approach the uniform limit as $\overline{\rho}\to \rho_{c}$ from both directions, strictly uniform solution $\rho=\rho_{c}=\overline{\rho}$ is incompatible with Eqs.(\ref{1d_Poisson}) and (\ref{combined}).
      
Near the critical density $\overline{\rho}=\rho_{c}$ explicit solutions to Eqs.(\ref{1d_Poisson}) and (\ref{combined}) can be given without resorting to models for the heat function $w(\rho)$.  Indeed, expanding the right-hand side of Eq.(\ref{combined}) to second order in $\rho-\rho_{c}$ we find 
\begin{equation}
\label{potential_near_rho_c}
\phi=-\frac{a}{2}(\rho-\rho_{c})^{2},~~a=\frac{1}{\rho_{c}^{3}} \Big\{ \frac{d}{d\rho}\left (\rho^{3}\frac{dw}{d\rho}\right )\Big\}_{\rho=\rho_{c}}>0.
\end{equation}
Adopting the following units, $[x]=(a\rho_{c}/8\pi G)^{1/2}$, $[\phi]=a\rho_{c}^{2}/2$, and $[\rho]=\rho_{c}$, brings the Poisson equation (\ref{1d_Poisson}) to the form
\begin{equation}
\label{critical_Poisson}
\phi''=1\pm \sqrt{-\phi}
\end{equation} 
whose solutions are subject to the boundary condition $\lim_{x\to \pm 0}\phi_{\pm}(x)=\mp g$.  Since $-\phi\ll1$ we limit ourselves to the $g\ll 1$ domain.  Then solutions for the potential and density in the form of power series about the origin are sufficient:
\begin{equation}
\label{power_series_potential}
\phi_{\pm}(x)=-g|x|+\frac{x^{2}}{2}\pm\frac{4\sqrt{g}}{15}|x|^{5/2}\mp\frac{1}{35\sqrt{g}}|x|^{7/2}+...
\end{equation} 
\begin{equation}
\label{power_series_density}
\rho_{\pm}(x)=1\pm\sqrt{g|x|}\mp\frac{1}{4\sqrt{g}}|x|^{3/2}+...
\end{equation}
These formulae hold within one period $[-g;g]$ and have to be periodically continued beyond it.  One can verify that $\overline{\rho}-1=\pm17g/30$, i.e. the difference between the average density and the critical density vanishes as $g\rightarrow 0$.  

\textit{Nonlinear gravitostatic waves.} - In an inertial reference frame traveling along the $x$-axis the steady solutions become traveling plane gravitostatic waves.  This term is introduced in Ref. \cite{Weinberg} to draw contrast with linear electrostatic (plasma) waves.  While linear gravitostatic waves are impossible, nonlinear ones can exists.  Their properties can be deduced from those of the steady flows.  Indeed, for a one-dimensional motion along the $x$ axis, $\textbf{v}=(v,0,0)$, we seek solutions for the density $\rho$, velocity $v$ and gravitational potential $\phi$ that depend only on $\xi=x-ut$ where $u$ is the velocity of the wave.  Then Eqs.(\ref{continuity}) and (\ref{2nd_law}) transform into       
\begin{equation}
\label{wave_continuity}
(-u\rho+\rho v)'=0
\end{equation}
\begin{equation}
\label{wave_Euler}
-uv'+vv'=-(w+\phi)'
\end{equation}
where now the prime refers to the derivative with respect to $\xi=x-ut$;  the  Poisson equation (\ref{1d_Poisson}) still formally holds.   

Integrating  Eq.(\ref{wave_continuity}) we find $-u\rho+\rho v=const=-u\overline{\rho}$ where the integration constant is fixed by the requirement of absence of average mass flux in the wave, $\overline{\rho v}=0$.  As a result one obtains
\begin{equation}
\label{integrated_wave_continuity}
\rho=\overline{\rho}\frac{u}{u-v}.
\end{equation} 
Thus the particles cannot travel faster than the wave, $v\leqslant u$, and their velocity changes sign at $\rho=\overline{\rho}$:  the particles are moving in the positive $x$-direction in the region where $\rho>\overline{\rho}$ and in the negative one if $\rho<\overline{\rho}$.  

Integrating Eq.(\ref{wave_Euler}) we arrive at the counterpart of the Bernoulli equation (\ref{Bernoulli})
\begin{equation}
\label{wave_Bernoulli}
\frac{(u-v)^{2}}{2}+w+\phi=const.
\end{equation}
If Eqs.(\ref{integrated_wave_continuity}) and (\ref{wave_Bernoulli}) are combined, we return to Eq.(\ref{combined}) with $j=\overline{\rho}u$.  Hereafter the results of the analysis of the steady flows carry over to the case of the gravitostatic waves via the substitutions $j\to \overline{\rho}u$ and $x\to \xi=x-ut$.  

We find it useful to restate them in a language better suited to traveling waves.  Indeed, the expression defining the critical density $\rho_{c}$ (\ref{critical_density}) can be rewritten as
\begin{equation}
\label{Mach}
M\equiv\frac{u}{s(\overline{\rho})}=\frac{\rho_{c}s(\rho_{c})}{\overline{\rho}s(\overline{\rho})}
\end{equation} 
where $s(\rho)=(\rho dw/d\rho)^{1/2}$ is the speed of sound and $M$ is a Mach number evaluated at average density $\overline{\rho}$.  We now observe that whether the left-hand side of Eq.(\ref{Mach}) is below or above unity is dictated by whether the critical density $\rho_{c}$ is below or above the average density $\overline{\rho}$, respectively.  Thus, if the motion is subsonic, $M<1$, the wave represents two coupled out of phase anharmonic oscillations of matter density and gravitational potential;  the hydrostatic limit corresponds to $M=0$.  On the other hand, if the motion is supersonic, $M>1$, the density-potential oscillations are in phase;  the zero-pressure limit corresponds to $M=\infty$.  In this particular case, according to Eq.(\ref{integrated_wave_continuity}), the particles responsible for the density singularity in Figure \ref{zero} travel with the velocity of the wave $v=u$ which allows them to "ride" the top of the potential.  Additionally, the amplitude of the gravitational field and the period of the wave (\ref{equalities}) will be given by 
\begin{equation}
\label{wave_equalities}
g=2u\sqrt{3\pi G\overline{\rho}},~~~\lambda=u \sqrt{\frac{3}{\pi G \overline{\rho}}}
\end{equation}
We observe that for $u$ fixed the gravitational field vanishes while the lattice period diverges as $\overline{\rho}\to 0$.         

\textit{Acknowledgements}. - I thank I. Shlosman for a valuable discussion that stimulated this work.

\end{document}